\documentclass[1p,12pt,authoryear]{elsarticle}

\usepackage{amssymb}
\usepackage{amsthm}
\usepackage{lineno}
\usepackage{graphicx}
\usepackage[ruled,vlined]{algorithm2e}

\usepackage{soul,xcolor,color}
\usepackage{setspace}
\usepackage{amsmath}
\usepackage{threeparttable} %For adding notes below tables
\graphicspath{{./Figures/}}
\usepackage{xr}
\usepackage{url}

\begin{document}

\begin{frontmatter}
\title{Machine learning enhanced data assimilation framework for multi-scale carbonate rock characterization}
\date{March 2025}

\author[1]{Zhenkai Bo \corref{cor1}}
\author[1]{Ahmed H. Elsheikh}
\author[1]{Hannah P. Menke}
\author[1]{Julien Maes}
\author[2]{Sebastian Geiger}
\author[3]{Muhammad Z. Kashim}
\author[3]{Zainol A. A. Bakar}
\author[1]{Kamaljit Singh}
\affiliation[1]{organization={Institute of GeoEnergy Engineering, Heriot-Watt University},
            city={Edinburgh},
            postcode={EH14 4AS}, 
            country={United Kingdom}}
\affiliation[2]{organization={Department of Geoscience \& Engineering, Technical University of Delft},
            city={Delft},
            country={The Netherlands}}
\affiliation[3]{organization={PETRONAS Research Sdn Bhd},
            city={Bangi, Selangor},
            country={Malaysia}}
\cortext[cor1]{Corresponding author.}
\begin{abstract}
Carbonate reservoirs offer significant capacity for subsurface carbon storage, oil production, underground hydrogen storage, geothermal energy, and groundwater flow. Accurate characterization of fluid flow behavior in these rocks is therefore critical for both resource recovery and emissions mitigation, yet it remains challenging due to their inherent heterogeneity. The wide range of carbonate pore-throat size distribution, spanning from nm to cm, hinders a comprehensive pore structure characterization with conventional single-scale X-ray computed tomography (micro-CT) images. Multi-scale imaging, which refers to acquiring CT images at both macro and micro resolutions, has emerged as a practical strategy to bridge this gap. In practice, nm-scale CT imaging requires the physical extraction of mini-plugs from the macro core sample, while the selection of drilling locations remains largely subjective, lacking a quantitative framework for rigorous decision-making. Digital rock modeling can assist this decision-making process by predicting flow properties from macro-scale sample images. However, its computational cost remains prohibitive for routine use. To facilitate an efficient, digitalized sub-sampling decision-making workflow, we propose a machine learning-enhanced data assimilation framework that leverages experimental drainage relative permeability measurements to achieve efficient characterization of micro-scale structures. We train a dense neural network (DNN) as a proxy to a multi-scale pore network simulator and couple it with an ensemble smoother with multiple data assimilation (ESMDA) algorithm. The DNN-ESMDA framework simultaneously infers the $CO_2$-brine drainage relative permeability of microporosity phases with associated uncertainty estimation, revealing the relative importance of each rock phase and guiding future characterization. Our DNN-ESMDA framework achieves a significant computational speedup, reducing inference time from thousands of hours to seconds compared to conventional multi-scale numerical simulation. The machine learning-enhanced ESMDA framework therefore provides a practical approach for improving the multi-scale imaging workflow of carbonates.

\end{abstract}


\end{frontmatter}

%\linenumbers
\pagenumbering{arabic}

\section{Introduction}
Characterization of multiphase flow behavior in carbonate rocks plays a significant role in many geoenergy applications, such as carbon capture and storage (CCS) \citep{orivri2025opportunities,krevor2015capillary}, oil and gas production \citep{shi2025pore}, underground hydrogen storage \citep{rezaei2022relative}, geothermal energy \citep{goldscheider2010thermal}, and groundwater flow \citep{Fryar2021}. Core-flooding experiments with X-ray computed tomography (X-ray CT) on cm-scale core samples are conducted to measure relative permeability of various sandstone and carbonate rock samples \citep{ruprecht2014hysteretic}. Such image data, coupled with experimental observations, can be used for building a digital rock model of sandstone rock samples to predict relative permeability at different capillary numbers \citep{jackson2020representative}. This is not always the case for carbonate rocks \citep{wenck2021simulating}, as they have a pore size distribution across scales ($10^{-9}$ to $10^{-2} \, m$) \citep{foroughi2024incorporation,wang2022anchoring} and the corresponding CO$_2$ multiphase flow behavior is difficult to characterize.  

Multi-scale characterization has emerged as a framework for addressing this multi-scale complexity, with methods developing over nearly two decades to capture pore structures across length scales. In this context, early work developed dual network models to incorporate information of both unresolved micro-porous regions and vugs (resolved void spaces in low-resolution images), reconstructing multiscale structure of carbonates \citep{biswal2007stochastic}. With continued advancement in rock imaging techniques over the past two decades, increasingly complex and realistic multi-scale pore network models are built based on images at different resolutions \citep{jiang2013representation,bultreys2015multi}. In these multi-scale simulations, small-scale features, e.g., microporosity phases of carbonate rocks \citep{menke2022using,wang2022anchoring}, \textcolor{red}{are typically characterized from high-resolution X-ray computed tomography images (X-ray CT) extracted from samples, and their properties are used as numerical input into the macroscale simulation domain for high-fidelity prediction.} By assuming the sampled high-resolution images as the representative elementary volume (REV), interpolated micro-structure-property relationships from the images are used during multi-scale simulations. Two critical considerations affect the efficacy of the workflow. One is the segmentation criteria for various phases (resolved pore, microporosity, and solid phases), and the other is the sampling bias.\cite{wang2022anchoring} found that segmenting microporosity phases based on their capillary pressure behaviour will lead to more accurate saturation map prediction during drainage in a mm-scale Estaillades core sample. It is also routine for digital rock studies to perform sensitivity analysis on the threshold values used for segmenting pore phases \citep{jackson2022deep}. Sampling bias arises from treating high-resolution X-ray CT images as REV for heterogeneous rocks. In this case, computed results from digital rock simulation may be significantly different from those of experimental measurement and require systematic correction \citep{saxena2019rock}. This limitation stems from high-resolution images having a field of view (FoV) that is inherently limited to approximately 1/1000th of the rock sample, a direct consequence of the sample size restrictions of high-resolution X-ray CT imaging. From this perspective, sampled mini-plugs may fail to realistically represent microscale heterogeneity and propagate systematic error into the multi-scale simulation workflow. 

One straightforward solution is to sample more microscale information to set up better REV or interpolation relationships. The problem is that preparing samples for high-resolution X-ray CT imaging is destructive, and exhaustive imaging of the whole FoV of a macroscale ($> 10^{-1} \;m$) sample is infeasible. \cite{menke2022using} showcased the sampling process of using multi-scale imaging workflow to predict the relative permeability point value during the drainage process of a mm-scale Estaillades core sample. To obtain nm-scale images, $\mu m$-scale sub-samples have to be drilled from the mm-scale core sample, where the pore structure information surrounding the extracted sub-sample is inevitably lost. \textcolor{red}{From this perspective, sub-sampling of carbonate rocks is irreversible, and, to the best of our knowledge, there is no quantitative method to assess the potential uncertainty reduction achieved through manually designated regions for sub-sampling.} Therefore, instead of randomly sampling from the regions of interest (ROI), we need a tool to inform the sampling location that can lead to effective uncertainty reduction. In this circumstance, we propose the usage of the ensemble smoother with multiple data assimilation (ESMDA) algorithm \citep{emerick2013ensemble} to quantify the uncertainty of multi-scale models and inform further sampling to improve the representation of microscale heterogeneity.   

ESMDA is well-suited in the context of multi-scale simulation of porous materials for several reasons. First, there are a great number of parameters to be determined in multi-scale numerical modeling of porous materials. A manual trial-and-error regression is often required \citep{foroughi2024incorporation}, but an automatic calibration method might lead to better results. Second, because of the limited sampling as mentioned previously, there is still a lack of a validation method to reveal the uncertainty introduced by using sampled microscale properties as REV \citep{norris2024uncertainty}. ESMDA, as a variant of the ensemble Kalman filter \citep{emerick2013ensemble}, is recognized for solving high-dimensional inverse problems while being capable of quantifying the inherent uncertainty \citep{zhou2022deep}. Consequently, ESMDA provides a comprehensive solution that simultaneously performs high-dimensional regression to experimental observations and quantifies uncertainty from each microscale parameter input, informing potential further characterization. 

During ESMDA regression for parameters in multi-scale simulations, forward modeling (simulation) is performed for each ensemble member to calculate the difference between the current prediction and the observations. This requires hundreds to thousands of numerical simulations for each implementation. However, in contrast to some large-scale numerical simulations in hydrology or reservoir engineering fields, which can be finished in minutes, pore-scale numerical simulations are often more computationally expensive \citep{menke2022using}, impeding ESMDA application. With the rapid advancement of machine learning algorithms, many neural networks are trained to predict properties of porous media X-ray CT images \citep{tembely2021machine,wang2024lattice}, accelerating the implementation of conventional characterization workflow \citep{delpisheh2024leveraging}. From this perspective, it is reasonable to leverage machine learning as a surrogate model to replace computationally intensive pore-scale numerical simulations within the ESMDA framework, thereby making inverse modeling and uncertainty estimation feasible for multi-scale characterization of porous materials. Note that neural networks have been regularly combined with the ESMDA algorithm in many subsurface applications, such as achieving regression under non-Gaussian distribution assumption \citep{zhang2020using}, simultaneously inferring high-dimensional parameters \citep{zhou2022deep}, and characterizing fracture media \citep{chen2025deep}. To the best of our knowledge, this study represents the first attempt to quantitatively guide the sub-sampling process for carbonate multi-scale imaging using machine learning, addressing a methodological gap that has not been previously recognized or resolved.

In this study, a dense neural network (DNN) is trained to act as a proxy model to the Extensive Pore Modeling (xpm, \url{https://github.com/dp-69/xpm}) multi-scale pore network (PNM) simulator, establishing a DNN-ESMDA framework for fast inference and uncertainty estimation of micro-scale properties in multi-scale rocks. We use a case study with a cm-scale Malaysian carbonate core sample to demonstrate the efficacy and compatibility of the DNN-ESMDA framework. Specifically, potential computational time for ESMDA regression is reduced from thousands of hours to seconds, making the inference of microporosity phase relative permeability of carbonate rocks from the whole-core ($38\; mm$ diameter $\times \; 69\; mm$ length) experimental measurements feasible. Moreover, uncertainty estimation of ensembles reveals valuable insights into future characterization and highlights the promising potential of this framework for routine application in multi-scale carbonate rock modeling workflows. 

The overall structure of this paper is as follows. Section \ref{sec:methodology} introduces the Malaysian carbonate samples, the image processing methods, and the ESMDA algorithm implementation details. Section \ref{sec:results} presents the validation of the results after ESMDA regression. Section \ref{sec:discussion} provides a summary of the main findings from the results and the computational cost of the ESMDA algorithm. Lastly, the section \ref{sec:conclusion} presents the concluding remarks.

\section{Methodology}\label{sec:methodology}
This section first introduces the Malaysian carbonate sample with corresponding image processing, then details the ESMDA methodology and demonstrates its implementation through the case study that validates the framework. 

\subsection{Heterogeneous multi-scale carbonate rock sample description}
Relative permeability measurements are expensive and hard to perform for carbonate rock samples and their associated microporosity phases. To address this challenge, we use multi-scale imaging techniques where high-resolution images are sampled from core samples to define the property models of microporosity phases. These models are then used to set up multi-scale digital rock models from micro-CT images for relative permeability prediction \citep{ruspini2021multiscale,menke2022using}. For heterogeneous carbonate rocks, there is always a question of whether the current sampling is representative enough to inform the multi-scale digital model to predict the properties of different rock samples \citep{pak2016multiscale}. This question is more pronounced for macroscale ($> 10^{-1} \;m$) carbonate rocks where both $\mu m-$ and $mm-$ scale heterogeneity affect the fluid flow. For such heterogeneous systems, sampling before uncertainty assessment may lead to unrecoverable information loss; therefore, leveraging high-resolution images from analogous rock samples can provide valuable initial constraints. Under these circumstances, it is beneficial to set up the relative permeability model for each microporosity phase based on whole-core experimental measurements, then use it to validate the high-resolution images and confirm their representativeness. Moreover, unlike invasion percolation for capillary pressure simulation, relative permeability prediction needs multi-scale simulation, which is computationally expensive to act as the forward model during ESMDA implementation. In this case study, we couple ESMDA with machine learning to overcome the computational challenges and infer relative permeability from limited observations on a cm-scale carbonate core sample. 

\begin{figure}[h]
  \centering
    \includegraphics[width=0.9\textwidth]{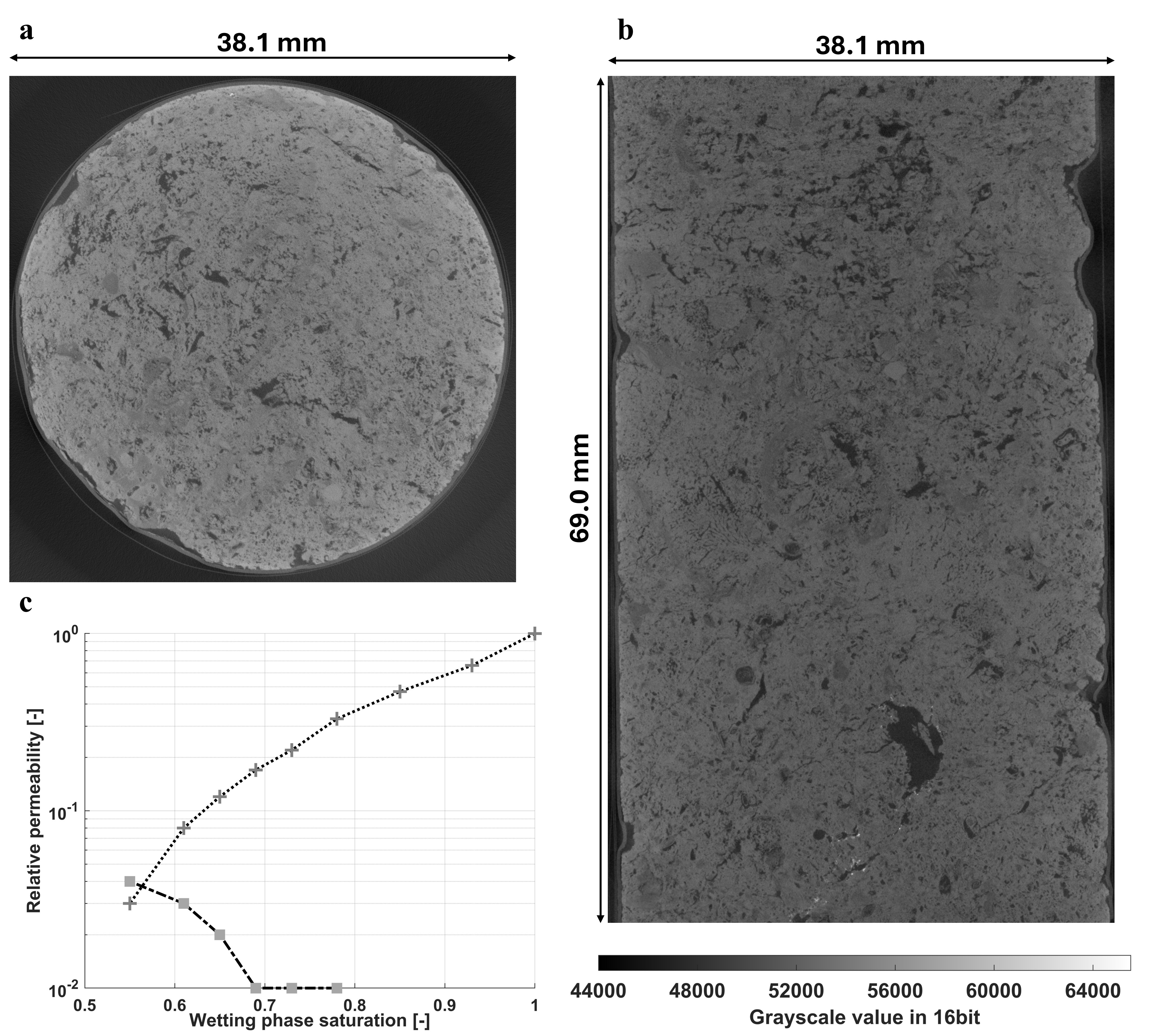}
    \caption{Raw micro-CT images of carbonate rock sample at $26.4\,\mathrm{\mu m}$ voxel size, black represents the resolved pore phase, gray represents the microporosity phase, and white represents the solid phase; (a) Circular cross section; (b) Internal structure along the core sample length; (c) Experimental measurement of steady-state $CO_2$-brine drainage relative permeability curves. }
    \label{fig:Rock_sample_description}
\end{figure}

\begin{figure}[h]
  \centering
    \includegraphics[width=1.0\textwidth]{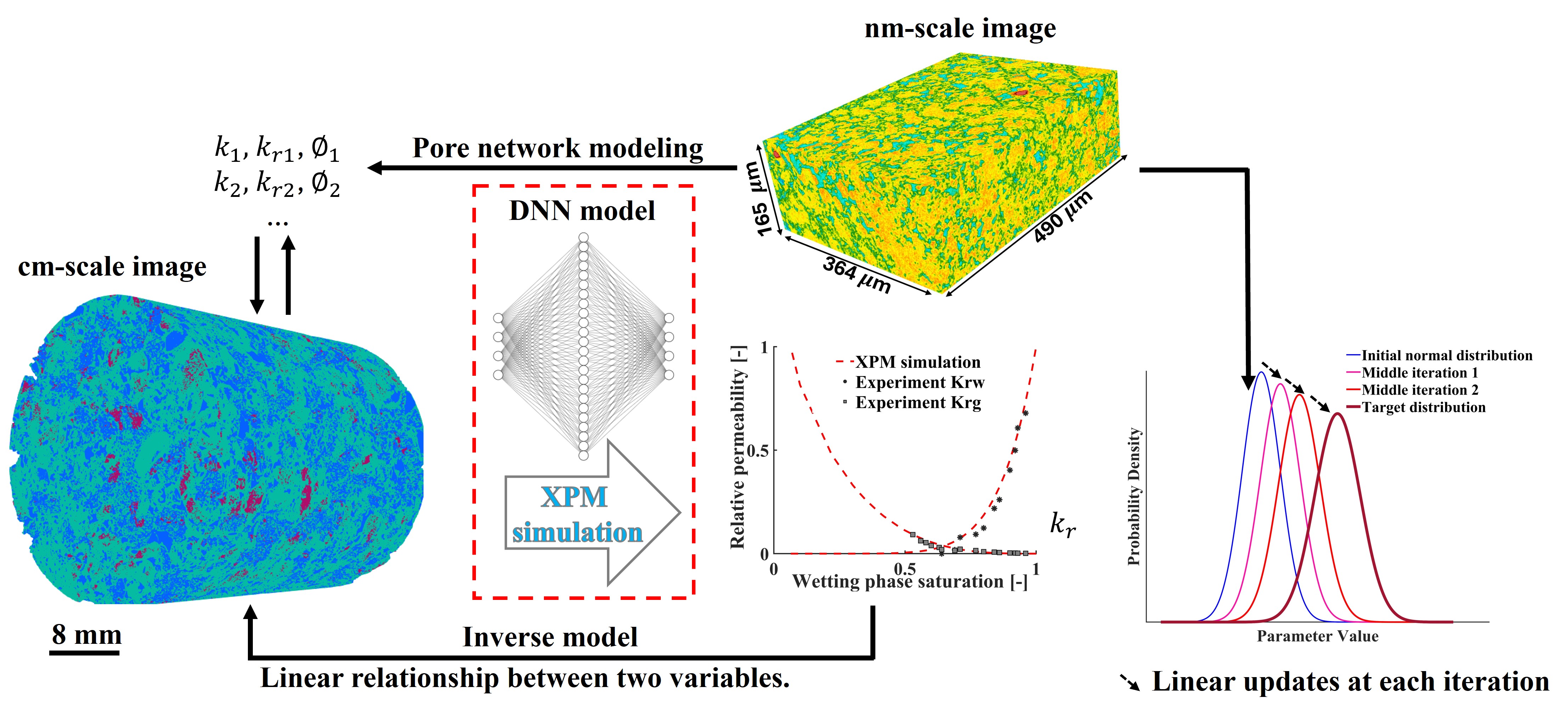}
    \caption{Schematic illustration of the general workflow of dense neural network (DNN) enabled data assimilation.}
    \label{fig:M1-47_multiscale_ESMDA_workflow}
\end{figure}

Our sample ($38\; mm$ diameter $\times \; 69\; mm$ length) is a Miocene bioclastic packstone from Central Luconia, deposited in a shallow marine environment, and consists of benthic foraminifera, coral reef, bryozoa, bivalves, and algae as shown in Figure \ref{fig:Rock_sample_description} (a) and (b). The experimental observations on this core sample are generally limited, consisting only of low-resolution micro-CT images, basic petrophysical properties (porosity and permeability), and experimentally measured drainage relative permeability curves (Figure \ref{fig:Rock_sample_description}). To compensate for these data scarcities and construct a digital rock model, we supplement our core-specific measurements with auxiliary data from analogous samples from the same formation, specifically a capillary pressure curve and a set of high-resolution images. Figure \ref{fig:M1-47_multiscale_ESMDA_workflow} presents a general workflow of our dense neural network (DNN) enabled data assimilation, where nm-scale images ($156 \times 364 \times 490 \; \mu m$) provide prior distribution of permeability, relative permeability, and capillary pressure curves. The following sections present the method to set up a digital rock model from cm-scale low-resolution micro-CT images and detail the implementation of a dense neural network-ESMDA framework for the inference of relative permeability models of each microporosity phase.

\subsection{Image processing and segmentation}
\begin{figure}[htp!]
  \centering
    \includegraphics[width=1.0\textwidth]{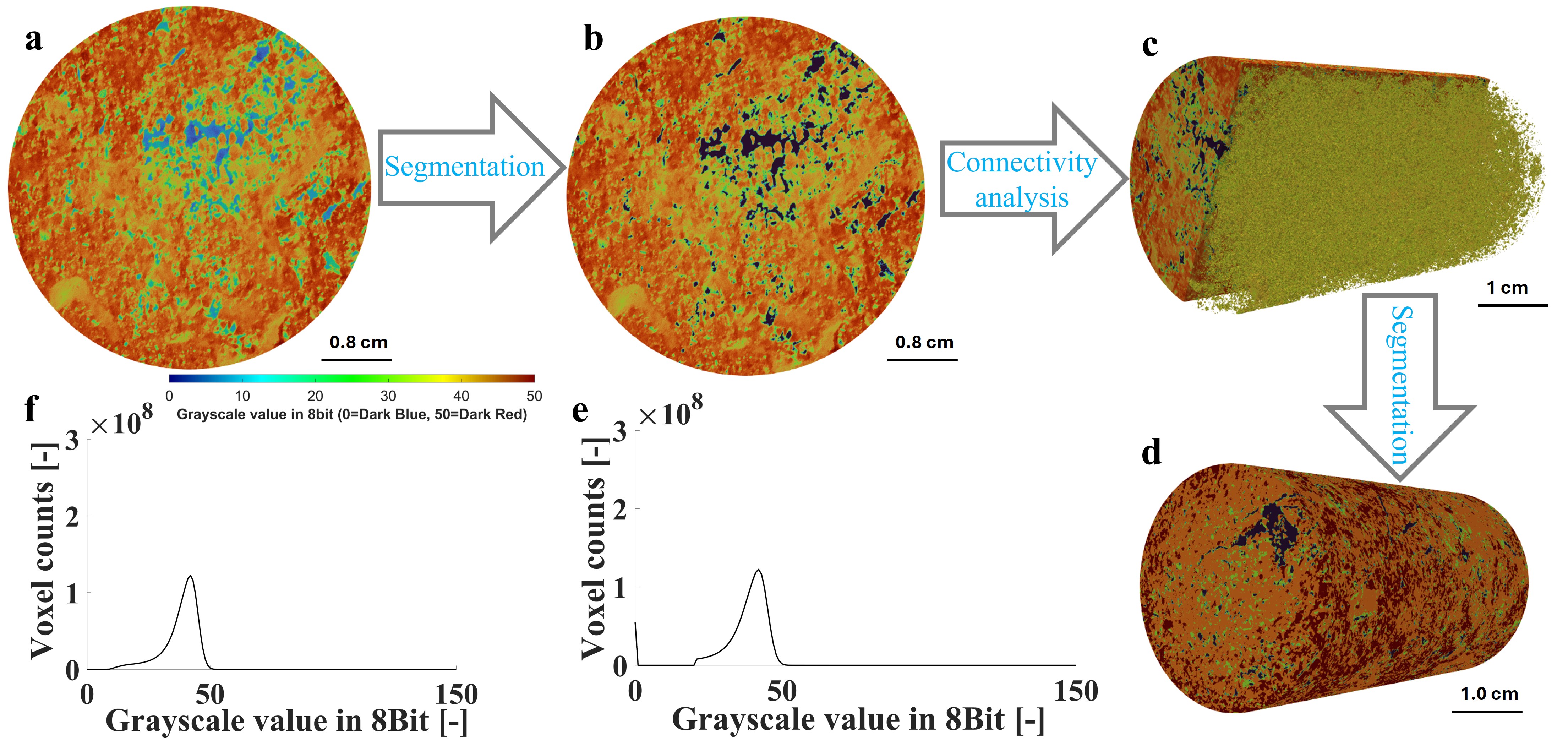}
    \caption{Schematic illustration of segmentation workflow for a dry-scan cm-scale carbonate rock sample from Malaysia at $26.4\,\mathrm{\mu m}$ voxel size; (a) Filtered dry-scan images, dark red represents the highest density, light red to green represent the microporosity regions, and the blue represent the resolved pore spaces; (b) Dry scan images with resolved pore spaces segmented as black color; (c) Two-end connectivity analysis showing the flow path provided by one of the grayscale value microporosity region; (d) Segmented images with three microporosity phases, presented as light blue, light green, and orange; (e) Grayscale value histogram of dry-scan images with resolved pore spaces segmented (volume fraction of resolved pore is 3.8\%) ; (f) Grayscale value histogram of dry-scan images. }
    \label{fig:Case_study_MalaysianCarbonate_seg_workflow}
\end{figure}

The raw X-ray CT image of the cm-scale carbonate core sample ($1000\times1000\times1801$ with a voxel size of $26.4\,\mathrm{\mu m}$, Figure \ref{fig:Rock_sample_description}) is first non-local mean filtered with an open-source Python script from \cite{spurin2024python}, shown as Figure \ref{fig:Case_study_MalaysianCarbonate_seg_workflow} a. After filtering, the resolved pore space can be segmented with visual thresholding, as in Figure \ref{fig:Case_study_MalaysianCarbonate_seg_workflow} b and e. 

During the drainage process of multi-scale PNM simulation, the non-wetting phase will first invade the connected resolved pore spaces and the microporosity phases with lower entry capillary pressure values. Then follow the microporosity phases with higher entry capillary pressure, resulting in greater non-wetting phase relative permeability at higher capillary pressure until irreducible wetting phase saturation is reached. Given the low resolution of the images, we assume microporosity regions are connected, and the entry capillary pressure value of a voxel increases with its grayscale value. Accordingly, the microporosity regions can be segmented based on the two-end connectivity, as this would be a direct indicator of the flow path for both wetting and nonwetting phases during multi-phase simulation. The segmentation of microporosity regions follows the scenario below:

\begin{itemize}

\item  From the lowest microporosity grayscale value $I_{start}$ where $ I_{start}>I_{resolved\_pore}$, create a binary mask including all resolved pores and all microporosity regions with grayscale $\leq I_{start}$, then test for percolation from one side of the image to the opposite side and calculate the fraction of resolved pore voxels that are part of the percolating flow path. If no percolation occurs, increment $I_{start}$ by 1 and repeat this procedure until finding the first microporosity grayscale value that enables percolation, marked as $I_{microporosity\_i}$.

\item Follow the previous procedure to find the $I_{microporosity\_{i+1}}$ which connects 99\% of resolved pore voxels, and the $I_{microporosity\_{i+2}}$ which connects 99.99\% of resolved pore voxels. 

\item Treat $I_{microporosity\_i}$, $I_{microporosity\_{i+1}}$, and $I_{microporosity\_{i+2}}$ as thresholds to segment the microporosity regions into 3 phases. Any grayscale values above $I_{microporosity\_{i+2}}$ are segmented as solid phase.

\end{itemize}

There are several reasons for using the thresholding method as above to segment the cm-scale images. The first reason is that the voxel size ($26.4\,\mathrm{\mu m}$) is not good enough to accurately capture the phase interfaces, considering the partial volume effects \citep{johns1993nondestructive}. Under these conditions, more advanced segmentation algorithms such as watershed or machine learning-based methods are unlikely to offer significant improvement, since the fundamental ambiguity lies in the image spatial resolution rather than in the choice of segmentation algorithm. Second, as shown in Figure \ref{fig:Case_study_MalaysianCarbonate_seg_workflow}, phases surrounding resolved pores are mostly microporosity, meaning the segmentation uncertainty brought by thresholding will not affect the number of connected voxels in our digital rock model. Considering we are segmenting the solid phase based on the percentage of connected resolved pore voxels, and the properties of microporosity voxels will be validated against experimental measurement, as discussed in the following sections, each stage of the workflow carries its own degree of segmentation-induced uncertainty. We believe this approach effectively disperses such uncertainty across these stages rather than concentrating it at a single step, thereby minimizing its overall impact on the simulation results. Moreover, our segmentation is connectivity-based rather than mineral-type-based, as the rock sample is composed of nearly pure calcite ($>99\%$). Such a method is routinely used in many multi-scale carbonate modelling studies \citep{menke2022using,wang2022anchoring,ruspini2021multiscale} to assign the same petrophysical properties to microporosity voxels that have similar flow behaviors from experimental measurement, avoiding over-fitting to the experimental observation.

\subsection{Petrophysical properties}
To predict the relative permeability for the whole core images, we need to define the porosity, permeability, capillary pressure, and relative permeability for each microporosity phase. With the segmented images, We define the porosity and capillary pressure curve for each phase by performing gradient-based regression against experimental data, following the scenarios detailed below. Note that both algorithms ran in seconds on an Intel(R) Xeon(R) Gold 6430 64-core CPU.

\begin{algorithm}[H]
\caption{Gradient-based porosity regression}
Calculate the target porosity contributed by all the microporosity phases $\phi_{target} = (\phi_{exp}-\phi_{resolved})/(V_1+V_2+V_3)$. $V_1$, $V_2$, and $V_3$ are the volume fractions of microporosity phases 1, 2, and 3, respectively;

Set the porosity range for the middle phase $\phi_2$, lower range is $\max(0.2, \phi_{target} - 0.05)$ and upper range is $\min(0.6, \phi_{target} + 0.05)$;

Give the upper range below 0.6 to microporosity phase 1,\\ $\phi_{1,min},\phi_{1,max} \gets \phi_{2,max}+ (0.6-\phi_{2,max})$;

Give the lower range above 0.2 to microporosity phase 3,\\ $\phi_{3,min},\phi_{3,max} \gets \phi_{2,min}-(\phi_{2,min}-0.2)$;

\While{$error \geq 0.005$}{
    $\phi_{core} \gets \phi_{resolved} + V_1\times\phi_1+V_2\times\phi_2+V_3\times\phi_3$\;
    
    $error \gets |\phi_{core}- \phi_{exp}|$\;
    
    $\Delta\phi_i \gets - error\times lr\times \frac{V_i}{V_1+V_2+V_3}$ for $i = 1,2,3$\;
    
    $\phi_i \gets \phi_i + \Delta\phi_i$ for $i = 1,2,3$\;
}
\end{algorithm}

\begin{algorithm}[H]
\caption{Gradient-based capillary pressure regression}
Initialize Brooks-Corey capillary pressure function for each phase\;

Give the entry capillary pressure of experimental measurement to microporosity phase 1, $P_{e1} \gets P_{c}^{exp}(S_{w1})$\;

Apply an increasing gradient of 1000 Pa for each phase, $P_{ei} \gets P_{e1} + 1000\times(i-1)$ for $i=2,3$\;

Initialize lambda with initial guess, $\lambda_i \gets 2$ for $i=1,2,3$\;

$iteration \gets 0$\;

Initialize learning rates $lr_{P_e}$ and $lr_{\lambda}$;

\While{$iteration < 400000$}{
    $\Delta P_{ei} \gets P_{ei} \times 0.01$\;
    
    $\Delta P_c \gets P_{c}^{percolation}(S_w,P_{ei}+\Delta P_{ei},\lambda_i) - P_{c}^{exp}(S_w)$\;
    
    $P_{ei} \gets P_{ei}- \Delta P_c\times lr_{P_e}$ for $i=1,2,3$\;
    
    $\Delta \lambda_i \gets \lambda_i \times 0.01$\;
    
    $\Delta P_c \gets P_{c}^{percolation}(S_w,P_{ei},\lambda_i+\Delta\lambda_i) - P_{c}^{exp}(S_w)$\;
    
    $\lambda_i \gets \lambda_i- \Delta P_c\times lr_{\lambda}$ for $i=1,2,3$\;
    
    $iteration \gets iteration + 1$\;
}
\end{algorithm}

After the porosity and capillary pressure model regression, the porosity of the whole core from the simulation is 0.346, while the experimental result is 0.347. The fitted capillary pressure is compared against MICP data until the residual wetting phase saturation is reached at around 0.55, which was obtained during the core-flooding experiment, shown in Figure \ref{fig:Case_study_MalaysianCarbonate_Pc_regressionr_results}.

\begin{figure}[htp!]
  \centering
    \includegraphics[width=1.0\textwidth]{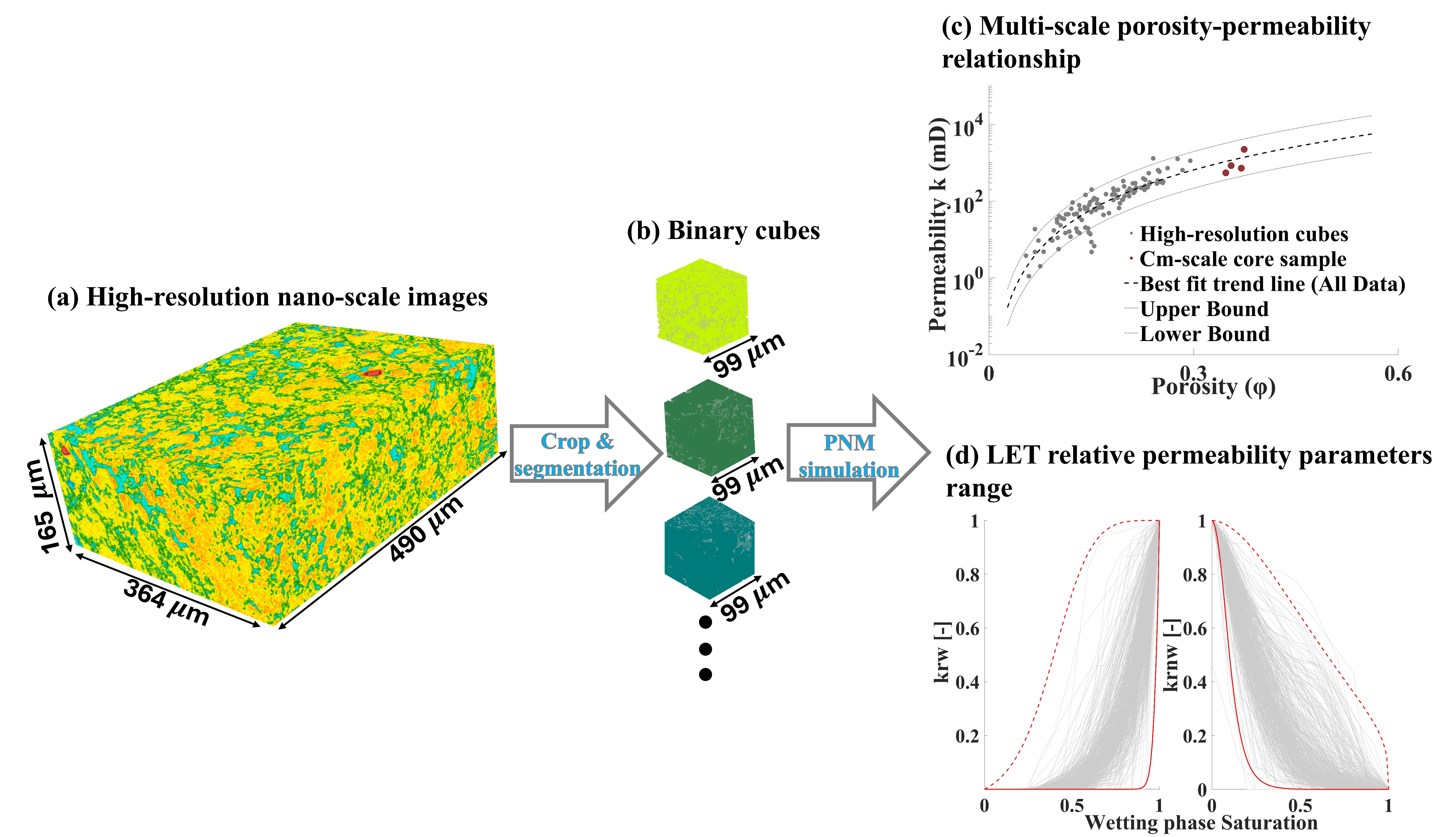}
    \caption{Workflow to define the permeability and Lomeland-Ebeltoft-Thomas (LET) relative permeability model parameter ranges; (a) high-resolution images at 165 nm from a analogous sample in the same formation; (b) the nano-scale images are cropped and segmented into 120 cube binary images; Run PNM simulation on these cubes and define: (c) permeability ($k=4.07e^4\varphi^{3.43}$) and (d) LET parameter ranges accordingly.}
    \label{fig:Case_study_MalaysianCarbonate_Poro-Perm_LET_range_define}
\end{figure}

We define the permeability and relative permeability models of microporosity phases in a data-driven way. As shown in Figure \ref{fig:Case_study_MalaysianCarbonate_Poro-Perm_LET_range_define}, we crop and segment the high-resolution images (ESRF France, $1000\times 2200\times 2970$ with voxel size of $165\,\mathrm{nm}$, \citep{yong2024multiscale}) from an analogous sample in the same formation into 120 binary sub-images with a dimension of $600^3$ (Figure \ref{fig:Case_study_MalaysianCarbonate_Poro-Perm_LET_range_define} b). PNM simulations with XPM (contact angle $45^o$) were performed on each of the sub-images to establish both porosity-permeability relationship (Figure \ref{fig:Case_study_MalaysianCarbonate_Poro-Perm_LET_range_define} c) and relative permeability range (Figure \ref{fig:Case_study_MalaysianCarbonate_Poro-Perm_LET_range_define} d). For permeability, we adjusted the pore-throat sizes of binary cube images to obtain a reasonable power law function fit (dashed black curve) between PNM simulation results and cm-scale core sample experimental measurements. This is done iteratively to ensure that the permeability bounds encompass both data. The established power law relationship and associated bounds are then used in multi-scale simulations, where permeability values are assigned to each microporosity phase based on its porosity. Accordingly, the permeability estimated from the multi-scale simulation is 544 mD, while the experimental measurement of the whole core is 547 mD.   

The relative permeability model used in this study is Lomeland-Ebeltoft-Thomas (LET) model \citep{lomeland2005new}, which is defined as 

\begin{equation}\label{eq:LET_equation_krw}
    \begin{aligned}
         k_{rw} =\frac{S_w^{L_w}}{S_w^{L_w}+E_w(1-S_w)^{T_w}}
    \end{aligned}
\end{equation}

\begin{equation}\label{eq:LET_equation_krnw}
    \begin{aligned}
         k_{rnw} =\frac{(1-S_w)^{L_n}}{(1-S_w)^{L_n}+E_n(S_w)^{T_n}}
    \end{aligned}
\end{equation}
where $L_w$, $E_w$, and $T_w$ are parameters for wetting phase relative permeability, and $L_n$, $E_n$, and $T_n$ are parameters for non-wetting phase relative permeability. As such, we can define the parameter prior distributions for the relative permeability models used in ESMDA by defining the distribution of these six parameters. LET model is used for relative permeability curve modeling because it matches better with data compared to other models, e.g., the Brooks-Corey type model \citep{bo2023impact}. As shown in Figure \ref{fig:Case_study_MalaysianCarbonate_Poro-Perm_LET_range_define} d, to estimate the prior ranges for the LET model parameters, we plot the PNM simulation results as light gray curves. By fitting the LET model to the upper and lower envelopes of these PNM results, we calculated the parameter ranges presented in Table \ref{Table LET_parameter_ranges}.

\begin{table}
\caption{Overview of LET parameter ranges from high-resolution nano-scale cube images}
\label{Table LET_parameter_ranges}
\begin{center}
\begin{tabular}{c c c c c c c}
\hline 
LET parameters & $L_w$ & $E_w$ & $T_w$  &  $L_n$ & $E_n$ & $T_n$ \\
\hline
Upper bound & 50 &3 &4 & 7.2 &30 &1.8 \\
Lower bound & 1 &3 & 1 & 0.35 & 1.5 &1.8 \\
\hline
\end{tabular}
\end{center}
\end{table}

\subsection{Ensemble smoother with multiple data assimilation (ESMDA)}
In this section, we provide a general introduction to ESMDA algorithm, where detailed implementation process, input, and output parameters can be found in section \ref{sec:methodology_DNN_ESMDA}. During EMSDA regression, the flow properties of each microporosity phase serve as input parameters for the multi-scale simulation on the digital rock model of carbonate rocks, and these parameters can be arranged into a vector $\mathbf{m}$ with dimension $N_m$. Before any experimental measurement, the probability distribution function of $\mathbf{m}$ is denoted as prior $\pi(\mathbf{m})$, which captures all the prior knowledge of the parameters $\mathbf{m}$ based on experience. The collected observations are stored in the vector $\mathbf{d}$. Given the observation error of $\mathbf{\epsilon}$ of the same dimension $N_d$, the experimental observation could be related to the forward multi-scale simulation model $\mathbf{g}$ using the following equation: 

\begin{equation}\label{eq:ESMDA_Observation}
    \begin{aligned}
         \mathbf{d} = \mathbf{g(m)}+\mathbf{\epsilon}
    \end{aligned}
\end{equation}
To set up a high-fidelity digital rock model for carbonate rocks, our goal is to update the prior distribution of $\pi(\mathbf{m})$ via assimilating experimental measurement $\mathbf{d}$, to obtain the posterior distribution of multi-scale simulation parameter $\pi(\mathbf{m}\mid\mathbf{d})$. This is achieved through Bayes' rule as 

\begin{equation}\label{eq:Bayes_rule}
    \begin{aligned}
         \pi(\mathbf{m}\mid\mathbf{d}) = \frac{\pi(\mathbf{m})\pi(\mathbf{d}\mid\mathbf{m})}{\pi(\mathbf{d})}, \; \pi(\mathbf{d}) = \int \pi(\mathbf{m})\pi(\mathbf{d}\mid\mathbf{m})d\mathbf{m}
    \end{aligned}
\end{equation}
where $\pi(\mathbf{d}\mid\mathbf{m})$ is the likelihood function, and $\pi(\mathbf{d})$ is the evidence that serves as a normalizing constant \citep{tarantola2005inverse}.

To possibly compute equation \ref{eq:Bayes_rule}, we use ESMDA \citep{emerick2013ensemble}, which is a variant of ensemble smoother (ES) or ensemble Kalman filter (EnKF). ES is initiated by drawing $N_e$ forecast (before assimilation) samples $\mathbf{M^{\textit{f}}}= [\mathbf{m}^{\textit{f}}_1,...\mathbf{m}^{\textit{f}}_{Ne}]$ from prior distribution $\pi(\mathbf{m})$, then linearly update them with 

\begin{equation}\label{eq:Ensemble_smoother}
    \begin{aligned}
         \mathbf{m}^a_j = \mathbf{m}^{\textit{f}}_j +\mathbf{C}^{\textit{f}}_{\mathbf{MD}}(\mathbf{C}^{\textit{f}}_{\mathbf{DD}}+\mathbf{C}_{\mathbf{D}})^{-1}[\mathbf{d}_{uc,j}-\mathbf{g}(\mathbf{m^\textit{f}_j})]
    \end{aligned}
\end{equation}
where $\mathbf{m}^a_j$ is the analysis ensembles conditioned on observation $\mathbf{d}$, $\mathbf{C}_D$ is the covariance matrix of observation error $\mathbf{\epsilon}$, $\mathbf{C}^{\textit{f}}_{\mathbf{DD}}$ is the auto-covariance of forward model predictions $\mathbf{D}^f=[\mathbf{m^\textit{f}_1},...,\mathbf{m^\textit{f}_{Ne}}]$, $\mathbf{g}(\mathbf{m})$ is the forward model prediction, $\mathbf{C}^f_{MD}$ is the cross-covariance between $\mathbf{M}^f$ and $\mathbf{D}^f$, lastly $d_{uc,j}$ is the perturbed observation sampling from Gaussian distribution $\mathcal{N}(\mathbf{d}, \mathbf{C}_D)$.

On the basis of ES, ESMDA employs multiple iterations of ES with an inflated covariance matrix to damp parameter updates at the early iterations. For iteration index $i = 1,...N_a$, our ESMDA implementation is 

\begin{equation}\label{eq:ESMDA_iterations}
    \begin{aligned}
         \mathbf{m}^{i+1}_j = \mathbf{m}^{\textit{i}}_j +\mathbf{C}^{\textit{i}}_{\mathbf{MD}}(\mathbf{C}^{\textit{i}}_{\mathbf{DD}}+\alpha_i\mathbf{C}_{\mathbf{D}})^{-1}[\mathbf{d}^i_{uc,j}-\mathbf{g}(\mathbf{m^\textit{i}_j})]
    \end{aligned}
\end{equation}
where $\sum^{N_a}_{i=1}\alpha_i = 1$ to ensure consistency with ES, and $\mathbf{d}^i_{uc,j}\sim \mathcal{N}(\mathbf{d},\alpha_i\mathbf{C_D})$. Under linear–Gaussian assumption, performing $N_a$ sequential assimilation steps, each using an inflated measurement error covariance (scaled by $N_a$), leads to the same overall update as applying the standard Ensemble Smoother (ES) update once \citep{emerick2013ensemble}.  

In this study, the input $\mathbf{m}^{\textit{i}}_j$, $\mathbf{g}(\mathbf{m^\textit{i}_j})$, and output $\mathbf{m}^{i+1}_j$ are set to be consistent with that of our multi-scale pore network modeling as discussed in the next sections. Thus, the non-linear parameter space of physical simulation can be explored linearly via ESMDA implementation.

\subsection{Multi-scale pore network modeling and machine learning-ESMDA framework}\label{sec:methodology_DNN_ESMDA}
\begin{figure}[htp!]
  \centering
    \includegraphics[width=1.0\textwidth]{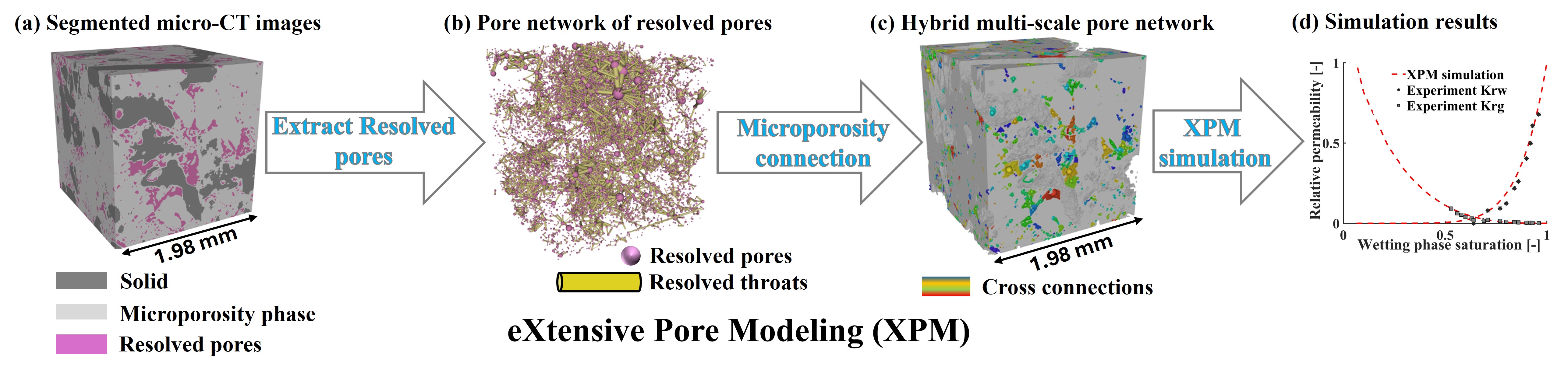}
    \caption{Schematic illustration of eXtensive Pore Modeling (XPM): (a) Multi-scale micro-CT images with resolved pores, microporosity phases, and solid phase segmented; (b) Extracted pore network from resolved pore regions with \textit{pnextract}; (c) Link microporosity voxels to their surrounding resolved pores, creating connection between microporosity phases and resolved pore regions, while different colors indicate which resolved pore each microporosity voxel is connected to; (d) Predict relative permeability from the multi-scale images by solving Stoke equation in resolved pore networks and Darcy equation in the Darcy cells (microporosity voxels).}
    \label{fig:XPM_workflow_demo}
\end{figure}

The well-fitted petrophysical properties are input into our in-house eXtensive Pore Modeling (XPM) (\url{https://github.com/dp-69/xpm}). As shown in Figure \ref{fig:XPM_workflow_demo}, XPM takes segmented images, extracts the resolved pore network structures, treats the unresolved microporosity voxels as Darcy cells, and assigns cross connections between the resolved pore network and the surrounding Darcy cells (microporosity voxels). During simulation, the Stokes equation is solved in resolved pore networks while the Darcy equation is solved in the Darcy cells. As such, after assigning microporosity phases properties into each Darcy cell and performing multi-scale PNM simulation, XPM can predict the quasi-static relative permeability of the micro-CT images and compare with experimental results. 

\begin{figure}[htp!]
  \centering
    \includegraphics[width=1.0\textwidth]{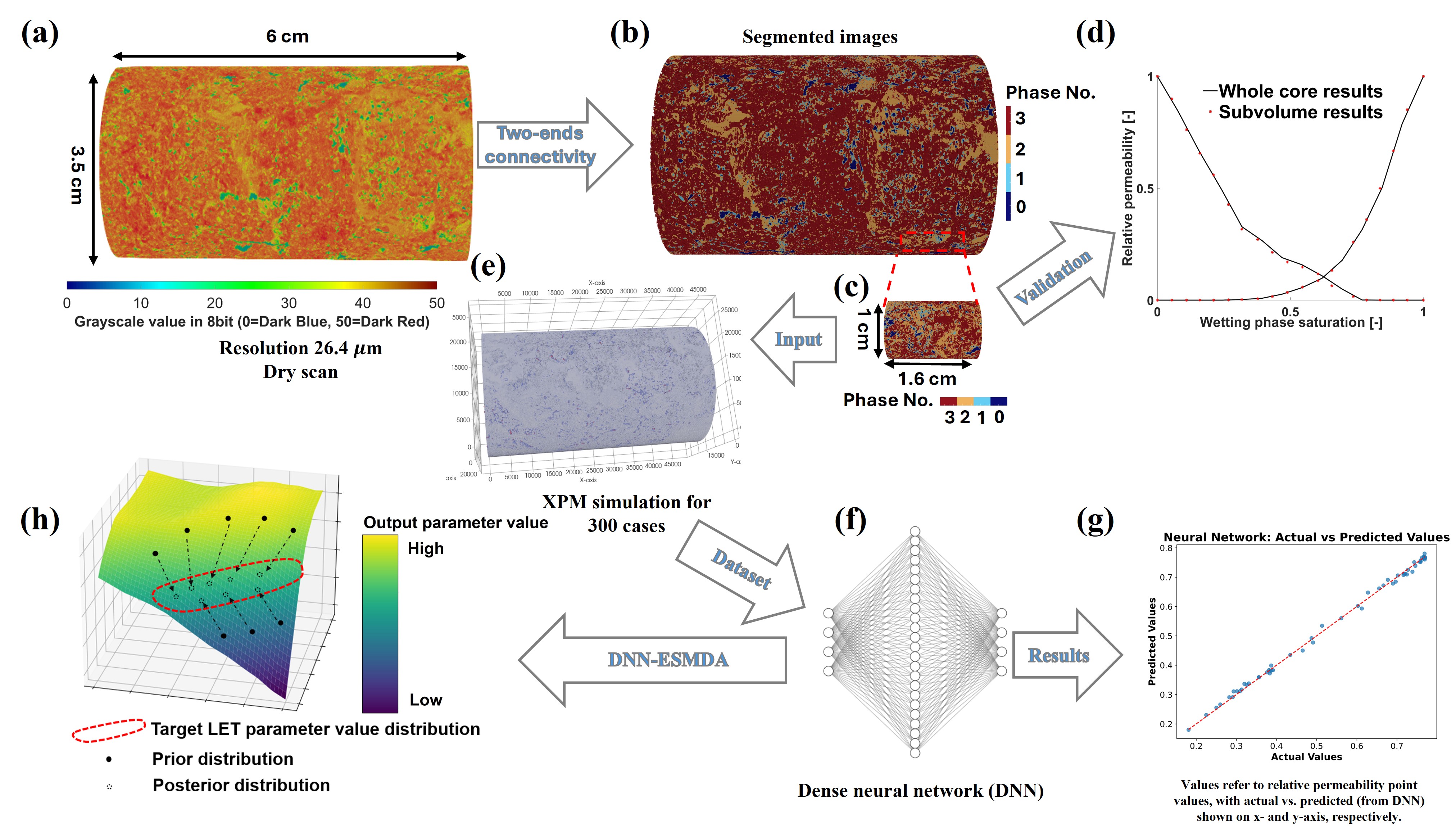}
    \caption{Schematic illustration of DNN-ESMDA workflow for fast inference of microporosity phase relative permeability of a cm-scale Malaysian carbonate sample: (a) Dry scan micro-CT images ; (b) Segmented images based on two-end connectivity, phase 0 as the resolved pore; (c) Subset the image into a subvolume for computational efficiency, phase 0 as the resolved pore; (d) Validate the subvolume simulation results with the whole core; (e) Use the subvolume as input image for 300 XPM simulation with Latin Hypercube sampling across the LET parameter space; (f) Train a three layer dense neural network (DNN) with XPM simulation results; (g) The training results of DNN, values refer to relative permeability point values, with actual vs. predicted (from DNN) shown on x- and y-axis, respectively; (h) Coupling DNN with ESMDA to establish a DNN-ESMDA framework for fast inference of microporosity phase relative permeability.}
    \label{fig:Casestudy_MalaysianCarboante_DNN-ESMDA_workflow_graphic_abstract}
\end{figure}

However, such a simulation would be computationally expensive to run hundreds of times, as the forward model in the ESMDA workflow. To achieve a fast inference of microporosity phases relative permeability from the prior distribution obtained from the high-resolution PNM simulations (Figure \ref{fig:Case_study_MalaysianCarbonate_Poro-Perm_LET_range_define}), we first subset the segmented images from Figure \ref{fig:Case_study_MalaysianCarbonate_seg_workflow} into a subvolume image with dimension of $400\times400\times600$ at voxel size of $26.4\,\mathrm{\mu m}$. Results show that this subvolume predicts similar relative permeability as the whole core images (Figure \ref{fig:Casestudy_MalaysianCarboante_DNN-ESMDA_workflow_graphic_abstract} d). Section \ref{app_sec_Malaysian_Carbonate_subvolume} provides details on how the subvolume image is selected. Using this subvolume image, we perform 300 XPM simulations with Latin Hypercube sampling across the LET parameter space. The generated simulation results provide training data for a DNN. The DNN is trained to map microporosity phase relative permeability LET parameters with point values on the predicted whole-core relative permeability curves, named $K_{r{1\_9}}$, $K_{r1\_13}$, and $K_{r2\_5}$ (see Figure \ref{app_fig:Real-Perm_point_value_DNN_output_illustration}). Two simple three-layer dense neural networks are used, where one takes the LET parameters $L_w$ and $T_w$ of two microporosity phases as input and predict the $K_{r1\_9}$ and $K_{r1\_13}$ as neural network output (four input and two output). The other neural network is trained to take LET parameters $L_n$ and $E_n$ of two microporosity phases as input and predict the $K_{r2\_5}$ as output (four input and one output). More details on the training and validation are presented in the appendix section \ref{appendix_sec:DNN_setting}, and Table \ref{Table_structure_input_output_DNNESMDA} provides an overview of the structure and training details of the DNN model. Coupling this DNN with ESMDA to establish a DNN-ESMDA framework facilitates the fast inference of microporosity phase relative permeability, as shown in Figure \ref{fig:Casestudy_MalaysianCarboante_DNN-ESMDA_workflow_graphic_abstract}.

\begin{table}
\caption{Overview of structure and training details of two dense neural networks.}
\label{Table_structure_input_output_DNNESMDA}
\begin{center}
\begin{tabular}{p {5 cm} p{4 cm}  p {4 cm}}
\hline 
Details & Neural network for wetting phase & Neural network for non-wetting phase \\
\hline
Input [-] &  $L_w$, $T_w$ & $L_n$, $E_n$ \\
Output [-] &  $K_{r1\_9}$, $K_{r1\_13}$ & $K_{r2\_5}$\\
Number of nodes in hidden layer [-] &  64 & 64\\
Activation function [-] & relu & relu \\
Learning rate [-] & 0.01 & 0.01\\
Batch size [-] &  32  & 16 \\
Training data percentage [\% of whole dataset] & 60 & 60 \\
Validation loss [-] & $7*10^{-4}$  & $8*10^{-5}$ \\ 
\hline
\end{tabular}
\end{center}
\end{table}

Our DNN-ESMDA for relative permeability inference is implemented following the scenario described below.

\begin{algorithm}[H]
\caption{DNN-ESMDA Iterative Update}
Set number of ESMDA iterations $N_a$ and corresponding inflation coefficients $\alpha_i$\;
Generate initial ensemble LET parameter values $\mathbf{m^{\textit{f}}_j}(j=1,...,N_e)$ of $L_w$, $T_w$, $L_n$, and $E_n$ from prior distribution $\mathbf{\pi(m)}$;

\For{$i \gets 1$ \KwTo $N_a$}{
    \For{$j \gets 1$ \KwTo $N_e$}{
        Input forecast LET parameters $\mathbf{m^{\textit{f}}_j}(j=1,...,N_e)$ into DNN to get the predicted relative permeability point values $\mathbf{g}(\mathbf{m^\textit{i}_j})$;
        
        Perturb observations $\mathbf{d}$ with inflated noise $\alpha_i \mathbf{C_D}$ to obtain $d^i_{uc,j}$\;
    }
    Compute cross covariance matrix $\mathbf{C}^{i}_{MD}$ and auto-covariance of predicted saturation $\mathbf{C}^{i}_{DD}$\;
    Update ensemble using Eq.~\eqref{eq:ESMDA_iterations} to obtain $\mathbf{m}^{i+1}_{j}(j=1,\dots,N_e)$\;
}
\end{algorithm}

\section{Results and discussion}\label{sec:results}

\subsection{DNN-ESMDA inference rsults}
Characterizing a cm-scale carbonate core sample needs a good understanding of both mm- and $\mu m-$ scale pore structure to capture the main flow mechanisms behind the quantitative experimental measurements, e.g., relative permeability curves. Consequently, sampling or sub-sampling the rock sample for high-resolution micro-CT scans are a necessity for validation. In this section, we present the results of DNN-ESMDA implementation for fast inference of microporosity phase relative permeability and demonstrate how the uncertainty assessment helps further sampling and characterization of the rock sample.

\begin{figure}[htp!]
  \centering
    \includegraphics[width=1.0\textwidth]{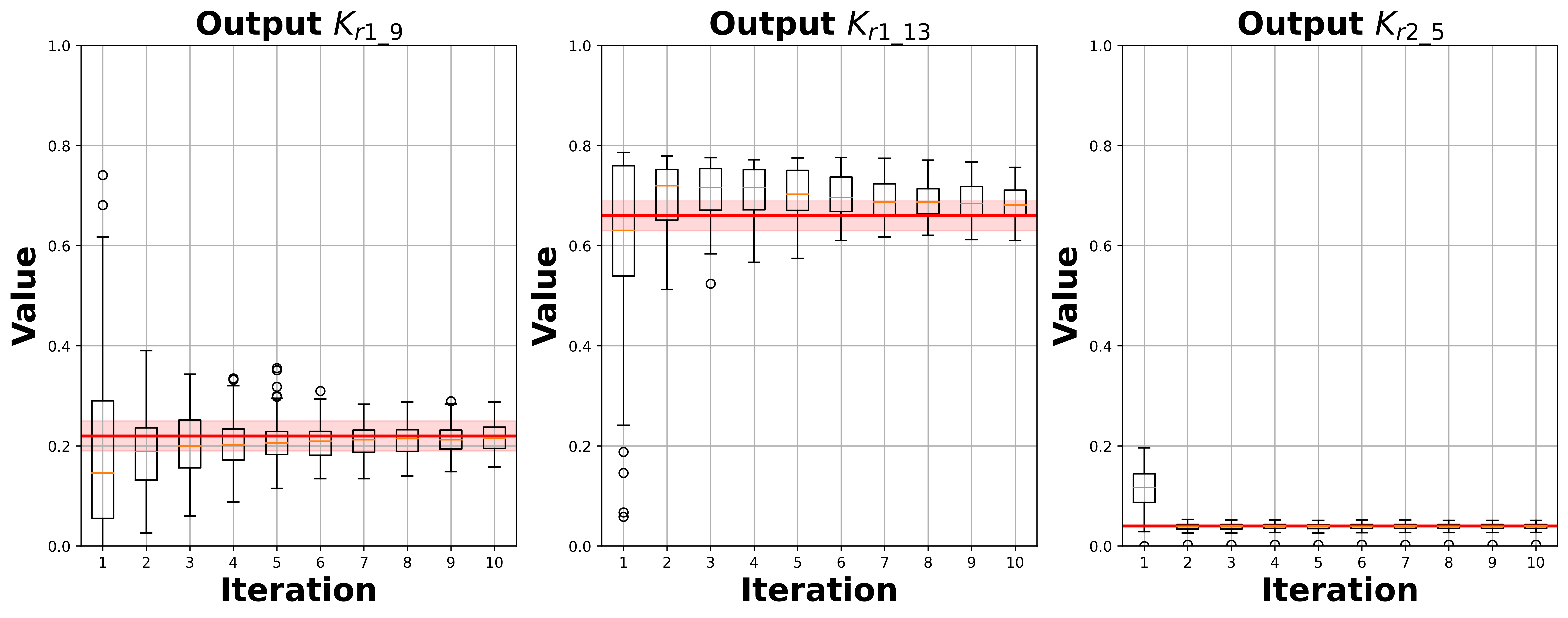}
    \caption{Box plots of ensemble prediction iteration history during DNN-ESMDA implementation for Malaysian carbonate rock sample; left: ensemble prediction iteration history of point value $K_{r1\_9}$ (wetting phase $9^{th}$ point value); middle: ensemble prediction iteration history of point value $K_{r1\_13}$ (wetting phase $13^{th}$ point value; right: ensemble prediction iteration history of point value $K_{r2\_5}$ (non-wetting phase $5^{th}$ point value, see appendix Figure \ref{app_fig:Real-Perm_point_value_DNN_output_illustration} for detailed definition); red line and shaded area are the target value and associated error, respectively.}
    \label{fig:Casestudy_Malaysian_carbonate_IterationHistory}
\end{figure}

The nm-scale high-resolution images from the analogous core sample provide us with a wide range of relative permeability curves. Starting from such a wide distribution, our DNN-ESMDA successfully shrinks the uncertainty of the prediction output effectively within a few iterations, shown in Figure \ref{fig:Casestudy_Malaysian_carbonate_IterationHistory}. Overall, our DNN-ESMDA framework can rapidly approach the target value (red line and shaded area) and keep ensembles from collapsing. This is demonstrated by the nearly constant ensemble uncertainty ranges in the last several iterations. Validation of the regression results is carried out by feeding the updated ensemble LET parameters back to XPM, then comparing the simulation results with experimental measurements. Figure \ref{app_fig:ESMDA_M1-47_validation} shows the validation results where our updated ensembles agree well with not only the three set target point values but also other experimental measurement points.

\begin{figure}[htp!]
  \centering
    \includegraphics[width=1.0\textwidth]{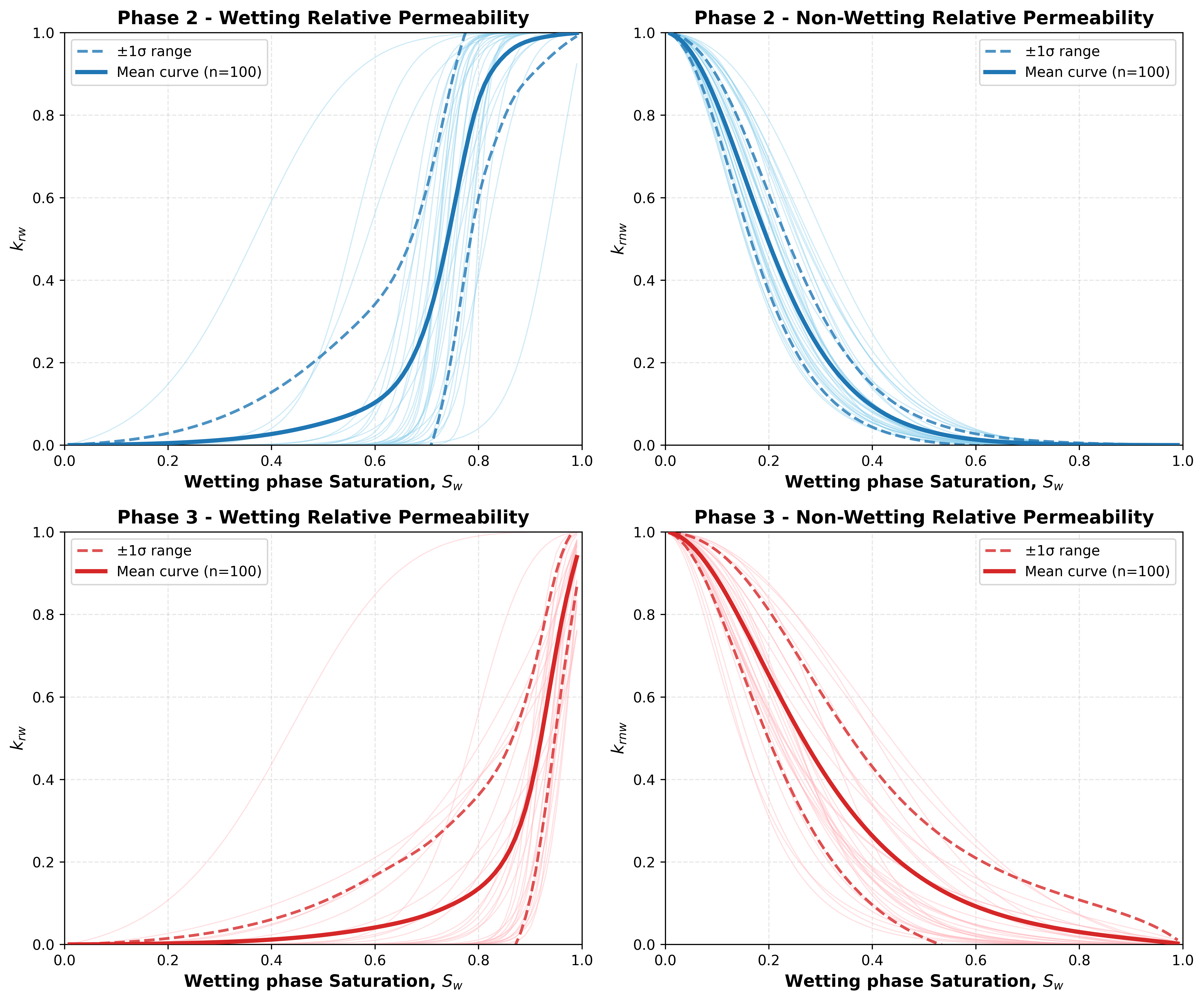}
    \caption{Updated relative permeability uncertainty of microporosity phase 2 (top row, see Figure \ref{fig:Casestudy_MalaysianCarboante_DNN-ESMDA_workflow_graphic_abstract} b and c) and phase 3 (bottom row, see Figure \ref{fig:Casestudy_MalaysianCarboante_DNN-ESMDA_workflow_graphic_abstract} b and c); solid lines mark the relative permeability curves from mean LET parameters out of the whole ensembles, while the dashed lines mark the one standard deviation range.}
    \label{fig:Casestudy_Malaysian_carbonate_ESMDA_ensemble_uncertainty}
\end{figure}

A further statistical analysis of the updated ensembles provides integrated information from multiple observation sources, e.g., low-resolution images and experimental relative permeability curves, encapsulated by ESMDA. Figure \ref{fig:Casestudy_Malaysian_carbonate_ESMDA_ensemble_uncertainty} presents the mean and one standard deviation range of the whole ensemble of microporosity phases 2 and 3 (Figure \ref{fig:Casestudy_MalaysianCarboante_DNN-ESMDA_workflow_graphic_abstract} c), respectively. Although we provide very limited information to the ESMDA algorithm, it can still deliver important information about the microporosity phases relative permeability uncertainty. From the left column of Figure \ref{fig:Casestudy_Malaysian_carbonate_ESMDA_ensemble_uncertainty}, despite the wide uncertainty range at the low relative permeability region of both microporosity phases wetting phase relative permeability, the P50 ($50^{th}$ percentile) of wetting phase relative permeability of phases 2 and 3 are 0.0014 and 0, respectively. In terms of non-wetting phase relative permeability as shown in the right column of Figure \ref{fig:Casestudy_Malaysian_carbonate_ESMDA_ensemble_uncertainty}, microporosity phase 2 has been well constrained into a narrow range, indicating that the input observations contain sufficient information to characterize these parameters, and they are critical for achieving good matches with the observations at hand. In contrast, the non-wetting phase relative permeability curves of phase 3 are less constrained and show a greater deviation range. From a core characterization perspective, these results entail that less information is captured regarding phase 3 compared to phase 2, and performing further imaging and core characterization on rock phases that are similar to phase 3 would improve understanding.

\subsection{Implications for multi-scale carbonate characterization} \label{sec:discussion}
Due to the multi-scale and multi-physics nature of porous materials characterization, experimental or computational observations from various origins are used to build a digital twin of samples that can reproduce or predict the physico-chemical processes observed during experimental measurements \citep{wang2022anchoring,foroughi2024incorporation,an2023inverse,wang2023large,fuchs2025generating}. Micro-structure properties of, e.g., heterogeneous carbonate rocks, are crucial parameters for physics-based simulation during such digital twin modeling.  While an explicit high-resolution X-ray CT imaging and direct multi-scale multi-physics simulation are not feasible \citep{fuchs2025generating}, our DNN-ESMDA framework can make computationally prohibitive regression practical and reveal the consistency of high-resolution images with experimental observation, providing valuable insights into future high-resolution imaging and characterization of porous materials. 

In our case study, there are no high-resolution X-ray CT images scanned for the microporosity phases of the cm-scale carbonate rock sample. Statistical distribution of microporosity phase relative permeability from a similar sample has to be used as an analogue. This situation is prevalent during the numerical modeling of porous materials, where analogue or empirical relationships instead of actual micro-structure properties are input into the simulation \citep{taghikhani2023electro,wang2022anchoring}. As shown in Figure \ref{fig:Casestudy_Malaysian_carbonate_ESMDA_ensemble_uncertainty}, implementing ESMDA regression will inform the consistency between the input statistics and the actual physical model. Any exception to the prior knowledge that is worth additional characterization will be identified. Moreover, coupling machine learning with ESMDA can potentially extend such identification capability. Input parameters of the numerical model are not the only variable that can affect the model predictions. X-ray CT image processing, e.g., noise filtering and segmentation, is another factor that would make a significant impact and is difficult to validate due to subjectivity. With machine learning, these processes can be parameterized, quantified, and fine-tuned through ESMDA \citep{niu2021geometrical}, achieving inference of micro-structure properties that conventional numerical simulation alone cannot provide.       

The computational cost of our DNN-ESMDA is affordable. Table \ref{Table Run_time_ESMDA} provides an overview of the total run time of DNN-ESMDA and corresponding XPM-ESMDA regression with an Intel(R) Xeon(R) Gold 6430 64-core CPU. Note that the run time of DNN-ESMDA refers to the DNN training time plus execution time of the $N_a$ ESMDA iterations, where the data processing time, e.g., image processing, is omitted here. Also, for certain numerical simulation problems, where neural networks or GPU can accelerate the solving process of partial differential equations (PDE) \citep{chen2026neural}, the implementation time difference between a DNN-ESMDA and a simulation-ESMDA might not be as distinct as presented here. Additionally, given the flexibility of numerical error control in traditional simulation methods, we recommend case-by-case consideration when choosing between DNN-ESMDA and simulation-based ESMDA.

\begin{table}
\caption{Comparison of total run times of DNN-ESMDA and XPM-ESMDA iterations in the case study}
\label{Table Run_time_ESMDA}
\begin{center}
\begin{tabular}{c c c c}
\hline 
Models & Ensemble size $N_e$ &Number of iteration $N_a$ & Total run time [s]\\
\hline
DNN-ESMDA & 100 & 10 & 346.66 \\ 
XPM-ESMDA & 100 & 10 & 2880000.00  \\ 
\hline
\end{tabular}
\end{center}
\end{table}

The Malaysian carbonate rock sample used in our study exhibits consistent nm-scale properties, as evidenced by the parameterizable porosity-permeability relationship and relative permeability obtained from nm-scale images. Given the promising potential of our proposed DNN-ESMDA framework, we encourage more multi-scale imaging studies on other types of carbonates to validate our workflow. When implemented with non-ideal datasets, e.g., parameters with non-Gaussian distributions, neural networks can be introduced to achieve similar regression performance and uncertainty assessment \citep{zhang2020using}. Moreover, the usage of any neural network as a replacement for numerical simulation will bring some degree of error into the system. Our measures to resolve this limitation are to perform a full numerical simulation with the final inferred ensemble input to ensure accuracy.

\section{Conclusions}\label{sec:conclusion}
\textcolor{red}{Comprehensive characterization of multi-scale carbonate rocks requires a deep understanding of both mm- and $\mu m$-scale pore structures, necessitating sub-sampling of rocks for high-resolution imaging. However, sub-sampling decisions in the current workflow are manual and lack of uncertainty quantification method} \citep{menke2022using,ruspini2021multiscale}. The ensemble smoother with multiple data assimilation (ESMDA) algorithm is a potential solution, while its implementation with multi-scale simulation would require excessive computational power.

In this study, we propose coupling neural networks with the ESMDA algorithm to infer CO$_2$ multiphase flow properties of microporosity phases in carbonate rocks. By using neural networks as computationally efficient proxies for expensive multi-scale multi-physics simulations, our framework enables both parameter regression and uncertainty estimation for carbonate sub-sampling problems that would otherwise be computationally prohibitive. In our demonstrative case study with a cm-scale Malaysian heterogeneous carbonate rock sample, we employ a dense neural network (DNN) as a proxy for our in-house Extensive Pore Modeling simulator (xpm, \url{https://github.com/dp-69/xpm}). The resulting DNN-ESMDA framework dramatically reduces the inference time for microporosity phase relative permeability from thousands of hours to seconds, while simultaneously quantifying the associated uncertainties for each phase. \textcolor{red}{Uncertainty analysis show distinct difference between two microporosity phases which are unperceivable from low-resolution images. Moreover, further comparison between prior (Figure} \ref{fig:Case_study_MalaysianCarbonate_Poro-Perm_LET_range_define}) and posterior statistics (Figure \ref{fig:Casestudy_Malaysian_carbonate_ESMDA_ensemble_uncertainty})  \textcolor{red}{demonstrates the capability of the DNN-ESMDA framework to assess the consistency between prior sub-sampling statistics and experimental observations, thereby guiding more confident and uncertainty-aware sub-sampling decisions for future multi-scale characterization of carbonate rocks}.

%Given its computational efficiency and promising potential, the machine learning-ESMDA framework can become part of routine characterization workflows for multi-scale porous materials across diverse applications after careful validation against broader datasets.

\section{Acknowledgments}
This work is funded by the Engineering and Physical Sciences Research Council's ECO-AI Project grant (reference number EP/Y006143/1), with additional financial support from the PETRONAS Centre of Excellence in Subsurface Engineering and Energy Transition (PACESET).

\section{Data availability}
The image data and DNN-ESMDA implementation source code can be found in \cite{bo_2026_20367421} and GitHub repository https://github.com/DigiPorFlow/AI-ESMDA4XPM.

%\appendix
%\renewcommand{\thesection}{S\arabic{section}}
%\setcounter{section}{0}

\bibliographystyle{elsarticle-harv}
\bibliography{Reference.bib}

@article{foroughi2024incorporation,
  title={Incorporation of sub-resolution porosity into two-phase flow models with a multiscale pore network for complex microporous rocks},
  author={Foroughi, Sajjad and Bijeljic, Branko and Gao, Ying and Blunt, Martin J},
  journal={Water Resources Research},
  volume={60},
  number={4},
  pages={e2023WR036393},
  year={2024},
  publisher={Wiley Online Library}
}

@article{menke2022using,
  title={Using nano-XRM and high-contrast imaging to inform micro-porosity permeability during Stokes--Brinkman single and two-phase flow simulations on micro-CT images},
  author={Menke, Hannah P and Gao, Ying and Linden, Sven and Andrew, Matthew G},
  journal={Frontiers in Water},
  volume={4},
  pages={935035},
  year={2022},
  publisher={Frontiers Media SA}
}

@article{bultreys2015multi,
  title={Multi-scale, micro-computed tomography-based pore network models to simulate drainage in heterogeneous rocks},
  author={Bultreys, Tom and Van Hoorebeke, Luc and Cnudde, Veerle},
  journal={Advances in Water resources},
  volume={78},
  pages={36--49},
  year={2015},
  publisher={Elsevier}
}

@article{wang2022anchoring,
  title={Anchoring multi-scale models to micron-scale imaging of multiphase flow in rocks},
  author={Wang, Shan and Ruspini, Leonardo C and {\O}ren, P{\aa}l-Eric and Van Offenwert, Stefanie and Bultreys, Tom},
  journal={Water Resources Research},
  volume={58},
  number={1},
  pages={e2021WR030870},
  year={2022},
  publisher={Wiley Online Library}
}

@article{ruspini2021multiscale,
  title={Multiscale digital rock analysis for complex rocks},
  author={Ruspini, LC and {\O}ren, PE and Berg, S and Masalmeh, S and Bultreys, Tom and Taberner, C and Sorop, T and Marcelis, F and Appel, M and Freeman, J and others},
  journal={Transport in Porous Media},
  volume={139},
  number={2},
  pages={301--325},
  year={2021},
  publisher={Springer}
}

@article{spurin2024python,
  title={Python workflow for segmenting multiphase flow in porous rocks},
  author={Spurin, Catherine and Ellman, Sharon and Sherburn, Dane and Bultreys, Tom and Tchelepi, Hamdi A},
  journal={Transport in Porous Media},
  pages={1--16},
  year={2024},
  publisher={Springer}
}

@article{emerick2013ensemble,
  title={Ensemble smoother with multiple data assimilation},
  author={Emerick, Alexandre A and Reynolds, Albert C},
  journal={Computers \& Geosciences},
  volume={55},
  pages={3--15},
  year={2013},
  publisher={Elsevier}
}

@article{zhou2022deep,
  title={Deep learning for simultaneous inference of hydraulic and transport properties},
  author={Zhou, Zitong and Zabaras, Nicholas and Tartakovsky, Daniel M},
  journal={Water Resources Research},
  volume={58},
  number={10},
  pages={e2021WR031438},
  year={2022},
  publisher={Wiley Online Library}
}

@inproceedings{lomeland2005new,
  title={A new versatile relative permeability correlation},
  author={Lomeland, Frode and Ebeltoft, Einar and Thomas, Wibeke Hammervold},
  booktitle={International symposium of the society of core analysts, Toronto, Canada},
  volume={112},
  pages={1--10},
  year={2005}
}

@article{an2023inverse,
  title={Inverse modeling of core flood experiments for predictive models of sandstone and carbonate rocks},
  author={An, Senyou and Wenck, Nele and Manoorkar, Sojwal and Berg, Steffen and Taberner, Conxita and Pini, Ronny and Krevor, Samuel},
  journal={Water Resources Research},
  volume={59},
  number={12},
  pages={e2023WR035526},
  year={2023},
  publisher={Wiley Online Library}
}

@book{tarantola2005inverse,
  title={Inverse problem theory and methods for model parameter estimation},
  author={Tarantola, Albert},
  year={2005},
  publisher={SIAM}
}

@article{krevor2015capillary,
  title={Capillary trapping for geologic carbon dioxide storage--From pore scale physics to field scale implications},
  author={Krevor, Samuel and Blunt, Martin J and Benson, Sally M and Pentland, Christopher H and Reynolds, Catriona and Al-Menhali, Ali and Niu, Ben},
  journal={International Journal of Greenhouse Gas Control},
  volume={40},
  pages={221--237},
  year={2015},
  publisher={Elsevier}
}

@article{norris2024uncertainty,
  title={Uncertainty quantification and propagation in lithium-ion battery electrodes using bayesian convolutional neural networks},
  author={Norris, Chance and Ayyaswamy, Abhinand and Vishnugopi, Bairav S and Martinez, Carianne and Roberts, Scott A and Mukherjee, Partha P},
  journal={Energy Storage Materials},
  volume={67},
  pages={103251},
  year={2024},
  publisher={Elsevier}
}

@article{tembely2021machine,
  title={Machine and deep learning for estimating the permeability of complex carbonate rock from X-ray micro-computed tomography},
  author={Tembely, Moussa and AlSumaiti, Ali M and Alameri, Waleed S},
  journal={Energy Reports},
  volume={7},
  pages={1460--1472},
  year={2021},
  publisher={Elsevier}
}

@article{wang2024lattice,
  title={Lattice Boltzmann prediction of CO2 and CH4 competitive adsorption in shale porous media accelerated by machine learning for CO2 sequestration and enhanced CH4 recovery},
  author={Wang, Han and Zhang, Mingshan and Xia, Xuanzhe and Tian, Zhenhua and Qin, Xiangjie and Cai, Jianchao},
  journal={Applied Energy},
  volume={370},
  pages={123638},
  year={2024},
  publisher={Elsevier}
}

@article{delpisheh2024leveraging,
  title={Leveraging machine learning in porous media},
  author={Delpisheh, Mostafa and Ebrahimpour, Benyamin and Fattahi, Abolfazl and Siavashi, Majid and Mir, Hamed and Mashhadimoslem, Hossein and Abdol, Mohammad Ali and Ghorbani, Mina and Shokri, Javad and Niblett, Daniel and others},
  journal={Journal of Materials Chemistry A},
  year={2024},
  publisher={Royal Society of Chemistry}
}

@article{wang2023large,
  title={Large-scale physically accurate modelling of real proton exchange membrane fuel cell with deep learning},
  author={Wang, Ying Da and Meyer, Quentin and Tang, Kunning and McClure, James E and White, Robin T and Kelly, Stephen T and Crawford, Matthew M and Iacoviello, Francesco and Brett, Dan JL and Shearing, Paul R and others},
  journal={Nature communications},
  volume={14},
  number={1},
  pages={745},
  year={2023},
  publisher={Nature Publishing Group UK London}
}

@article{fuchs2025generating,
  title={Generating multi-scale Li-ion battery cathode particles with radial grain architectures using stereological generative adversarial networks},
  author={Fuchs, Lukas and Furat, Orkun and Finegan, Donal P and Allen, Jeffery and Usseglio-Viretta, Francois LE and Ozdogru, Bertan and Weddle, Peter J and Smith, Kandler and Schmidt, Volker},
  journal={Communications Materials},
  volume={6},
  number={1},
  pages={4},
  year={2025},
  publisher={Nature Publishing Group UK London}
}

@article{taghikhani2023electro,
  title={Electro-chemo-mechanical finite-element model of single-crystal and polycrystalline NMC cathode particles embedded in an argyrodite solid electrolyte},
  author={Taghikhani, Kasra and Weddle, Peter J and Hoffman, Robert M and Berger, JR and Kee, Robert J},
  journal={Electrochimica Acta},
  volume={460},
  pages={142585},
  year={2023},
  publisher={Elsevier}
}

@article{niu2021geometrical,
  title={Geometrical-based generative adversarial network to enhance digital rock image quality},
  author={Niu, Yufu and Da Wang, Ying and Mostaghimi, Peyman and McClure, James E and Yin, Junqi and Armstrong, Ryan T},
  journal={Physical Review Applied},
  volume={15},
  number={6},
  pages={064033},
  year={2021},
  publisher={APS}
}

@article{orivri2025opportunities,
  title={Opportunities and challenges for geologic CO2 sequestration in carbonate reservoirs: A review},
  author={Orivri, Uzezi D and Chanda, Piyali and Johnson, Liz and Koehn, Lars W and Pollyea, Ryan M},
  journal={International Journal of Greenhouse Gas Control},
  volume={142},
  pages={104342},
  year={2025},
  publisher={Elsevier}
}

@article{ruprecht2014hysteretic,
  title={Hysteretic trapping and relative permeability of CO2 in sandstone at reservoir conditions},
  author={Ruprecht, Catherine and Pini, Ronny and Falta, Ronald and Benson, Sally and Murdoch, Lawrence},
  journal={International Journal of Greenhouse Gas Control},
  volume={27},
  pages={15--27},
  year={2014},
  publisher={Elsevier}
}

@article{jackson2020representative,
  title={Representative elementary volumes, hysteresis, and heterogeneity in multiphase flow from the pore to continuum scale},
  author={Jackson, Samuel J and Lin, Qingyang and Krevor, Sam},
  journal={Water Resources Research},
  volume={56},
  number={6},
  pages={e2019WR026396},
  year={2020},
  publisher={Wiley Online Library}
}

@article{wenck2021simulating,
  title={Simulating core floods in heterogeneous sandstone and carbonate rocks},
  author={Wenck, Nele and Jackson, Samuel J and Manoorkar, Sojwal and Muggeridge, Ann and Krevor, Samuel},
  journal={Water Resources Research},
  volume={57},
  number={9},
  pages={e2021WR030581},
  year={2021},
  publisher={Wiley Online Library}
}

@article{pak2016multiscale,
  title={Multiscale pore-network representation of heterogeneous carbonate rocks},
  author={Pak, Tannaz and Butler, Ian B and Geiger, Sebastian and Van Dijke, Marinus Ij and Jiang, Zeyun and Surmas, Rodrigo},
  journal={Water Resources Research},
  volume={52},
  number={7},
  pages={5433--5441},
  year={2016},
  publisher={Wiley Online Library}
}

@article{bo2023impact,
  title={Impact of experimentally measured relative permeability hysteresis on reservoir-scale performance of underground hydrogen storage (UHS)},
  author={Bo, Zhenkai and Boon, Maartje and Hajibeygi, Hadi and Hurter, Suzanne},
  journal={international journal of hydrogen energy},
  volume={48},
  number={36},
  pages={13527--13542},
  year={2023},
  publisher={Elsevier}
}

@article{rezaei2022relative,
  title={Relative permeability of hydrogen and aqueous brines in sandstones and carbonates at reservoir conditions},
  author={Rezaei, Amin and Hassanpouryouzband, Aliakbar and Molnar, Ian and Derikvand, Zeinab and Haszeldine, R Stuart and Edlmann, Katriona},
  journal={Geophysical Research Letters},
  volume={49},
  number={12},
  pages={e2022GL099433},
  year={2022},
  publisher={Wiley Online Library}
}

@article{shi2025pore,
  title={The pore-network-continuum modeling of two-phase flow properties for multiscale digital rocks},
  author={Shi, Bowen and Rong, Jianqi and Jiang, Han and Guo, Bo and Hassanizadeh, S Majid and Qin, Chao-Zhong},
  journal={Advances in Water Resources},
  pages={105138},
  year={2025},
  publisher={Elsevier}
}

@article{saxena2019rock,
  title={Rock properties from micro-CT images: Digital rock transforms for resolution, pore volume, and field of view},
  author={Saxena, Nishank and Hows, Amie and Hofmann, Ronny and Alpak, Faruk O and Dietderich, Jesse and Appel, Matthias and Freeman, Justin and De Jong, Hilko},
  journal={Advances in Water Resources},
  volume={134},
  pages={103419},
  year={2019},
  publisher={Elsevier}
}

@article{jackson2022deep,
  title={Deep learning of multiresolution x-ray micro-computed-tomography images for multiscale modeling},
  author={Jackson, Samuel J and Niu, Yufu and Manoorkar, Sojwal and Mostaghimi, Peyman and Armstrong, Ryan T},
  journal={Physical Review Applied},
  volume={17},
  number={5},
  pages={054046},
  year={2022},
  publisher={APS}
}

@article{johns1993nondestructive,
  title={Nondestructive measurements of fracture aperture in crystalline rock cores using X ray computed tomography},
  author={Johns, Robert A and Steude, John S and Castanier, Louis M and Roberts, Paul V},
  journal={Journal of Geophysical Research: Solid Earth},
  volume={98},
  number={B2},
  pages={1889--1900},
  year={1993},
  publisher={Wiley Online Library}
}

@article{chen2026neural,
  title={Neural Physics: Using AI Libraries to Develop Physics-Based Solvers for Incompressible Computational Fluid Dynamics},
  author={Chen, Boyang and Heaney, Claire E and Pain, Christopher C},
  journal={Computers \& Fluids},
  pages={106981},
  year={2026},
  publisher={Elsevier}
}

@article{zhang2020using,
  title={Using deep learning to improve ensemble smoother: Applications to subsurface characterization},
  author={Zhang, Jiangjiang and Zheng, Qiang and Wu, Laosheng and Zeng, Lingzao},
  journal={Water Resources Research},
  volume={56},
  number={12},
  pages={e2020WR027399},
  year={2020},
  publisher={Wiley Online Library}
}

@article{chen2025deep,
  title={Deep learning-based inversion framework for fractured media characterization by assimilating hydraulic tomography and thermal tracer tomography data: Numerical and field study},
  author={Chen, Cihai and Deng, Yaping and Qian, Jiazhong and Ma, Haichun and Ma, Lei and Wu, Jichun and Wu, Hui},
  journal={Engineering Geology},
  volume={350},
  pages={107998},
  year={2025},
  publisher={Elsevier}
}

@inproceedings{yong2024multiscale,
  title={Multiscale Upscaling Study for CO2 Storage in Carbonate Rocks Using Machine Learning and Multiscale Imaging},
  author={Yong, Wen Pin and Menke, Hannah and Maes, Julien and Geiger, Sebastian and Bakar, Zainol Affendi Abu and Lewis, Helen and Buckman, Jim and Bonnin, Anne and Singh, Kamaljit},
  booktitle={Offshore Technology Conference Asia},
  pages={D021S007R002},
  year={2024},
  organization={OTC}
}

@article{goldscheider2010thermal,
  title={Thermal water resources in carbonate rock aquifers},
  author={Goldscheider, Nico and M{\'a}dl-Sz{\H{o}}nyi, Judit and Er{\H{o}}ss, Anita and Schill, Eva},
  journal={Hydrogeology Journal},
  volume={18},
  number={6},
  pages={1303--1318},
  year={2010},
  publisher={Springer}
}

@incollection{Fryar2021,
  author    = {Fryar, Alan E.},
  title     = {Groundwater of carbonate aquifers},
  booktitle = {Global Groundwater: Source, Scarcity, Sustainability, Security and Solutions},
  editor    = {Mukherjee, Abhijit and Scanlon, Bridget R. and Aureli, Alice and Langan, Simon and Guo, Huaming and McKenzie, Andrew A.},
  publisher = {Elsevier},
  address   = {Amsterdam},
  year      = {2021},
  pages     = {23--34},
  doi       = {10.1016/B978-0-12-818172-0.00002-5}
}

@article{biswal2007stochastic,
  title={Stochastic multiscale model for carbonate rocks},
  author={Biswal, B and {\O}ren, P-E and Held, RJ and Bakke, S and Hilfer, R},
  journal={Physical Review E—Statistical, Nonlinear, and Soft Matter Physics},
  volume={75},
  number={6},
  pages={061303},
  year={2007},
  publisher={APS}
}

@article{jiang2013representation,
  title={Representation of multiscale heterogeneity via multiscale pore networks},
  author={Jiang, Zeyun and Van Dijke, MIJ and Sorbie, Kenneth Stuart and Couples, Gary Douglas},
  journal={Water resources research},
  volume={49},
  number={9},
  pages={5437--5449},
  year={2013},
  publisher={Wiley Online Library}
}

@misc{bo_2026_20367421,
  author       = {Bo, Zhenkai (Josh)},
  title        = {{Machine learning enhanced data assimilation framework 
                   for multi-scale carbonate rock characterization}},
  year         = {2026},
  month        = may,
  publisher    = {Zenodo},
  doi          = {https://doi.org/10.5281/zenodo.20367421},
  note         = {Version v1}
}

\end{document}